**Phonon-blocked junction refrigerators for cryogenic quantum devices**

# Phonon-blocked junction refrigerators for cryogenic quantum devices

E. Mykkänen[1], J. S. Lehtinen[1], A. Ronzani[1], A. Kemppinen[1], A. Alkurdi[2], P.-O. Chapuis[2], M. Prunnila[1]

[1]VTT Technical Research Centre of Finland, Espoo, Finland, email: mika.prunnila@vtt.fi
[2]FCETHIL UMR5008, Univ Lyon, CNRS, INSA-Lyon, Université Claude Bernard Lyon 1, F-69621, Villeurbanne

*Abstract*— Refrigeration is an important enabler for quantum technology. The very low energy of the fundamental excitations typically utilized in quantum technology devices and systems requires temperature well below 1 K. Expensive cryostats are utilized in reaching sub-1 K regime and solid-state cooling solutions would revolutionize the field. New electronic micro-coolers based on phonon-blocked semiconductor-superconductor junctions could provide a viable route to such miniaturization. Here, we investigate the performance limits of these junction refrigerators.

## I. INTRODUCTION

Quantum devices are typically operated in massive and expensive cryoliquid-based refrigerators capable to reach sub-100 mK temperature. Typically these devices only require their miniaturized active elements to reach such low temperature. Compact solid-state cooling solutions could significantly decrease costs, complexity, maintenance and physical space required to reach low temperatures. Integrated on-chip electronic coolers and miniaturized cooled platforms [1,2] solutions would enable e.g. wider-spread and cost-effective refrigeration for quantum information processing [3,4], single-photon detection [5] and spaceborne detectors [6].

In Ref. [7] we demonstrated a promising novel approach for cooling at sub-kelvin temperatures: A semiconductor - superconductor (Sm-S) tunnel junction, where thermionic cooling arises from the voltage controlled evacuation of the hotter electrons through energy-gapped superconducting leads and acoustic phonon transmission bottleneck at the junction limits the parasitic phonon thermal conductance. The latter is the key innovation compared to previous works of superconducting tunnel junction coolers, where electron-phonon coupling in lateral cold fingers limits the operation [1].

The operation range of superconducting coolers scales with the critical temperature of the superconductor, $T_c$. In analogy with conventional thermoelectric coolers, it may be expanded by cascading stages with superconductors with different $T_c$ by using the 3D configuration introduced in Ref. [7]. Such cascade is depicted also in Fig. 1(a). In this communication, we investigate the operation of single stages of Sm-S junction coolers in two cases: when the phonon blocking is enhanced with constrictions and when thermionic cooling power is increased with the transparency of the junction up to the limit defined by higher-order tunneling processes.

## II. OPERATION PRINCIPLES AND MODELLING

Fig. 1(b) shows a conceptual image of a single phonon-blocked Sm-S tunnel junction cooling element. The operation principle is depicted in Figs. 1(c) and 1(d) with following four elements [7]: *(i)* weak electron-phonon coupling in the superconductor disconnects electron and phonon heat conduction channels, *(ii)* phonon branch heat transport is suppressed by thermal resistance $R_{\mathrm{ph}}$ (boundary resistance at the tunnel junction, $R_{\mathrm{PTB}}$, and thermal resistance of superconducting lead, $R_{\mathrm{lead}}$, in series), *(iii)* electron heat channel is suppressed by the superconductor energy gap $\Delta$, and *(iv)* the electron cooling is enabled by quasiparticle filtering/thermionic tunneling by the same energy gap.

The cooling power is a function of tunnel junction resistance, $R_T$, and leakage in the energy gap, $\gamma = R_T/R_{\mathrm{gap}}$, where $\gamma$ is the (Dynes) leakage parameter and $R_{\mathrm{gap}}$ is the electric resistance of the junction at zero voltage [8]. A simplified equation for cooling power at optimal voltage bias $V_{\mathrm{opt}} = (\Delta - 0.66 k_B T)/e$ can be written as [7]

$$P_{\mathrm{cool}}(T_N) = \frac{\Delta^2}{e^2 R_T} 0.59 \left(\frac{k_B T_N}{\Delta}\right)^{3/2} - \frac{1}{2}\frac{V_{\mathrm{opt}}^2}{R_{\mathrm{gap}}}, \qquad (1)$$

where $e$ is the elementary charge, $k_B$ is the Boltzmann constant and $T_N$ is the temperature of semiconductor [9]. Equation (1) is valid when temperature is well below critical temperature.

The detrimental phonon heat leak through a thermal resistance (for example $R_{PTB}$ or $R_{\mathrm{lead}}$) can be described by

$$P_{\mathrm{ph}}(T_N, T_0) = \frac{1}{\alpha n}(T_N^n - T_0^n). \qquad (2)$$

It is related to thermal resistance as $R(T) = \alpha T^{-n+1}$ when $T_N = T_0 = T$. Here $T_0$ is the temperature of the previous stage. Thermal resistance pre-factor $\alpha$ and power $n$ depend on the mechanism of the phonon transport. Temperature $T_N$ in the simulations is determined by setting $P_{\mathrm{ph}} = P_{\mathrm{cool}}$.

For planar $R_{\mathrm{PTB}}$ and 3D $R_{\mathrm{lead}}$ $n = 4$ [10,8] and for ballistic 1D channel $n = 2$ [11]. In Ref. [7] an experimental value for $R_{\mathrm{PTB}}$ in a geometry similar to Fig. 2(a) was obtained and ballistic 1D case (thermal conductance quantum limit) was investigated theoretically. Here we model also diffusive quasi 1D case relevant for nanowire constrictions [Fig. 2(b)]. The heat transport in constrictions is estimated with the phonon Boltzmann transport equation (BTE) under the relaxation time approximation solved by the Discrete Ordinate Method

(DOM). This allows observing the impact of the geometry and properties of the boundary on the transport properties, in particular the transition between ballistic limit and the Casimir regime, which is an effective diffusion regime due to reduction of phonon free path by interaction with the diffusely-scattering boundaries of the nanowires. However, it does not account for the reduction of the numbers of transversal phonon modes. The BTE under DOM simulations give $n = 3$ in (2).

As the dominant phonon wavelength (at the temperatures of interest) is several orders of magnitudes larger than the Fermi wavelength, the phonon channel becomes one-dimensional earlier than electron channel when the later size is reduced [12]. This enables efficient heat removal by electrons while detrimental phonon thermal leaks are strongly suppressed. The upper limit of phonon thermal resistance is set by a one thermal conductance quantum divided by the transmission probability.

### III. RESULTS AND DISCUSSION

We investigate three different scenarios: *(i)* the phonon flow limited by $R_{\text{PTB}}$ [Fig. 2(a)], *(ii)* the phonon flow limited by $R_{\text{PTB}}$ and nanowire constrictions at superconducting leads [Fig. 2(b)], and *(iii)* thermal conductance quantum limit. Fis. 2(c) and 2(d) show relative cooling as a function of (higher stage) temperature at the three scenarios for aluminum and vanadium. These two materials were chosen since they have been used previously in superconducting electron coolers [1,7,13]. Fig 2(c) also shows experimental data from Ref. [7], which is obtained from a similar prototype platform as in Fig. 1(e). The simulated results of the figures are calculated with realistic parameters $R_{PTB}$=22 K$^4$cm$^2$/W [8], $\gamma = 10^{-3}$ [14] and $R_T A$= 100 Ωμm$^2$, where $A$ is tunnel junction area. In the BTE nanowire simulations in Figs. 2(c) and 2(d), we have assumed aluminum cylinders with width-to-height ratio of 10 (width 50 nm) so that the total area of constriction wires is 1/3 of tunnel junction area.

Figure 2(c) shows that the chosen simulation values for resistance and leakage almost maximize the cooling effect in aluminum junctions without any need of constrictions and predict highly improved performance compared to the measured data of Ref. [7]. In contrast, vanadium would benefit from the constrictions. Figures 2(e) and 2(f) show the overall operation range at the classical ($n = 4$ in (2)) and conductance quantum ($n = 2$) limits as a function of temperature and relevant parameters and give guidelines for designing phonon-blocked cooler junctions.

As can be seen from (1), the cooling power can be enhanced by decreasing $R_T$. However, as the resistance decreases, junctions become more transparent and second-order tunneling process, Andreev reflection, may become relevant. The Andreev reflection tunneling rate is inversely proportional to the number of Andreev conduction channels, $N = A/A_{ch}$, where $A$ ($A_{ch}$) is the area of the junction (area of a single channel). In the limit of opaque junctions, $A_{ch}$ relates to other tunnel junction parameters as

$$\gamma = R_K A_{\text{ch}}/(4R_T A), \quad (3)$$

where $R_K = h/e^2$ is the resistance quantum ($h$ is Planck's constant). In Ref. [15] an experimental value for the Andreev channel area was found, $A_{\text{ch}}$ =30 nm$^2$. This value was about ten times larger than the one that is predicted by theory.

The Andreev tunneling limit (based on Ref. [15] experimental results) is shown in Figs. 3(a) and 3(b) when the total phonon resistance is 220 K$^4$cm$^2$/W, corresponding to a situation where the phonon flow is suppressed by constrictions in addition to the phonon thermal boundary resistance. The Figures can be used to identify the optimal operation range for highly-transparent tunnel junctions and guide fabrication efforts. Figs. 3(a) and 3(b) show that very good performance can be obtained even when Andreev reflection is taken into account.

### IV. SUMMARY

We have studied methods of enhancing the performance of phonon-blocked Sm-S tunnel junction coolers stages. Introduction of nanowire constrictions yields significant performance gain and extends the operation of the cooler to higher temperatures. Performance can be further enhanced by increasing the transparency of the tunnel junctions up to the point where higher order tunneling processes appear. As temperature difference larger than 50% per stage can be reached, our approach allows highly-attractive scalable and fully electronic refrigerator technology that can be used for quantum circuit applications in the 1 K - 100 mK temperature range with vanadium/aluminium stages. Even lower base temperature could be possible with a stage based on lower-gap superconductors such as titanium.

#### ACKNOWLEDGMENT

This project was financially supported by H2020 programme projects EFINED (766853) and MOS-QUITO (688539) and Academy of Finland projects QuMOS (288907, 287768) and ETHEC (322580). This work was performed as part of the Academy of Finland Centre of Excellence program (312294).

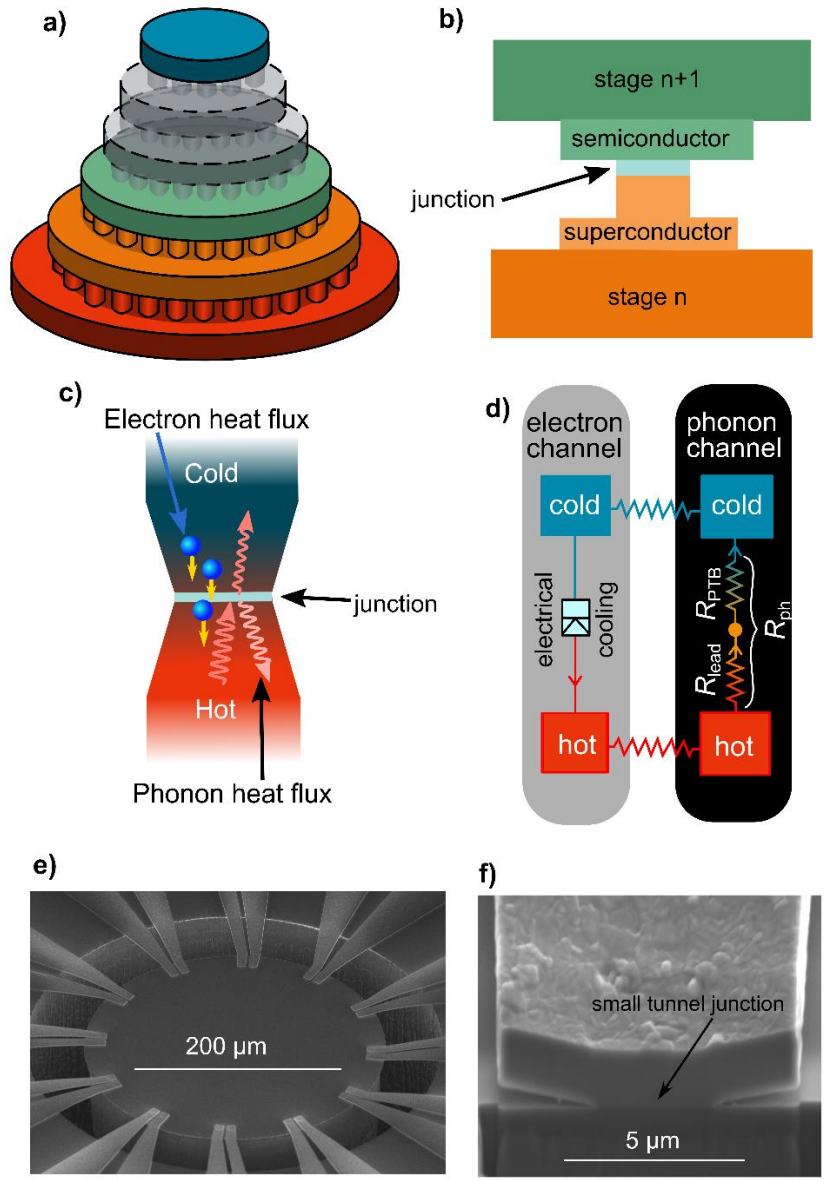

Fig. 1. (a) Conceptual image of a cascaded cooler structure (top: cold, bottom: hot), (b) implementation of a cooling element in the cascade, and (c) electron and phonon heat fluxes in a tunnel junction. (d) Heat transfer equivalent circuit. Here $R_{\text{lead}}$ corresponds to the thermal resistance of the superconducting lead, $R_{\text{PTB}}$ the thermal resistance of the junction and $R_{\text{ph}}$ their sum. (e), (f) Scanning electron micrographs of a cooler prototype studied in Ref. [8] to verify the phonon blocked operation. (f) Cross-section of a tunnel junction between the middle Si pillar and one superconducting Al lead. See also Ref. [8].

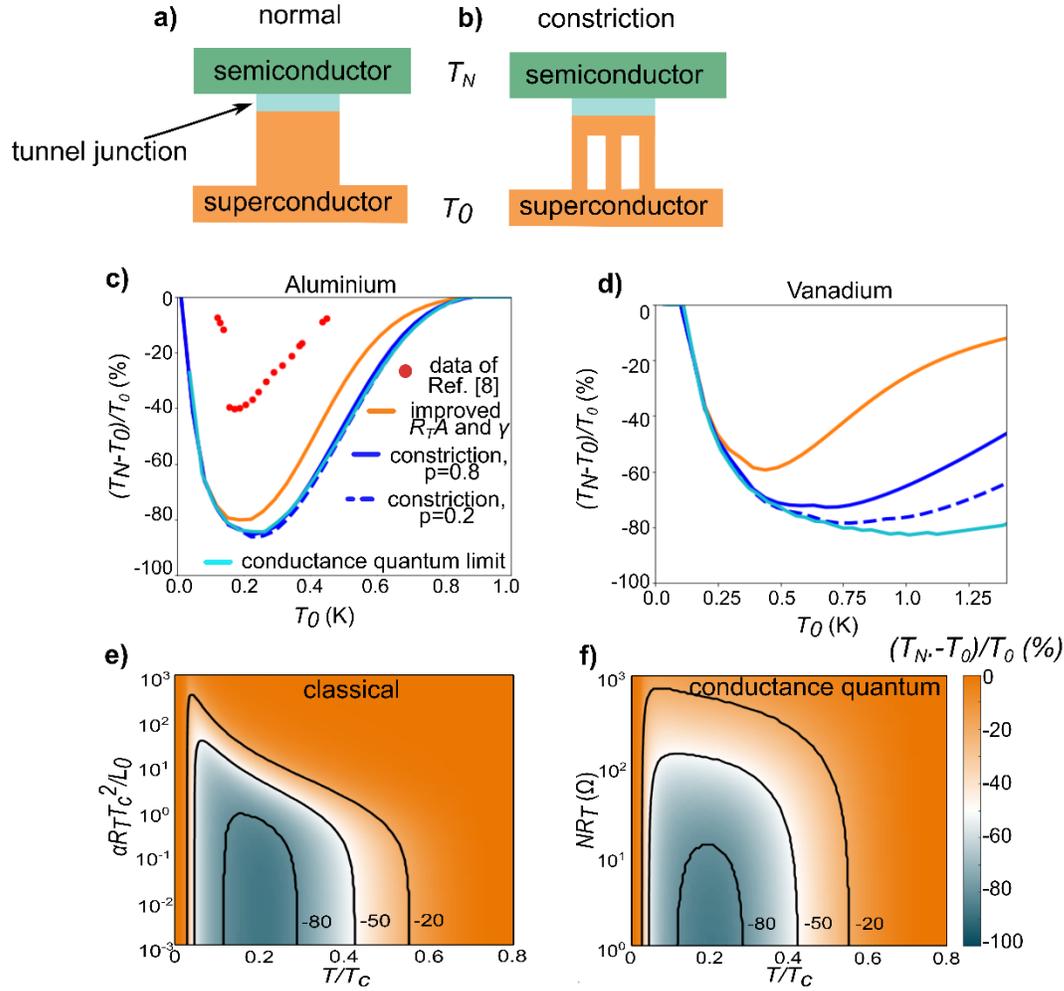

Fig. 2. (a)-(b) Different approaches for phonon filtering in superconductor: (a) Phonons are only filtered at the junction, and (b) nanowire constrictions are introduced. (c) Measurement data of Ref [8] (red dots). (c),(d) Simulations with tunnel junction with improved characteristic junction resistance, $R_T A$, and leakage parameter, $\gamma$ (orange line), nanowire constrictions with two different typical boundary reflection specularity parameters, $p$ (blue lines) and constriction that is limited by 10 conductance quantum (cyan line). The data is shown as percentual cooling between the previous (hot) stage with temperature, $T_0$ and subsequent cold stage with temperature $T_N$. (e) Simulated cooling with traditional phonon thermal boundary resistance and lead resistance as a function of normalized temperature $T/T_c$, and product of electrical cooling and phonon blocking parameters (see (1) and (2)). Here $L_0$ is Lorenz number and $T_c$ critical temperature of the superconductor. (f) Simulated cooling when phonon channel is limited by limited number of thermal conductance quanta, $N$.

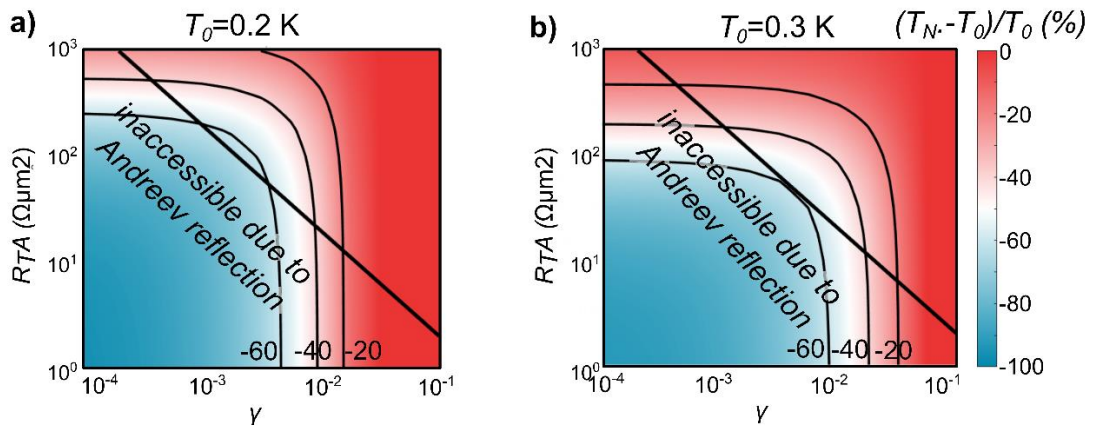

Fig. 3. Cooling at starting temperatures of 0.2 K and 0.3 K for aluminium when constrictions are introduced. Solid thick black line indicates the Andreev tunneling limit (see text).